\documentclass[twocolumn]{jpsj2} 
\title{Electronic states and pairing symmetry
in the two-dimensional 16 band d-p model for iron-based superconductor}

\author{Yuki \textsc{Yanagi}$^{1}$\thanks{E-mail address: yanagi@phys.sc.niigata-u.ac.jp}, Youichi \textsc{Yamakawa}$^{1}$, and Yoshiaki \textsc{\=Ono}$^{1,2}$}

\inst{$^{1}$Department of Physics, Niigata University, Ikarashi, Niigata
950-2181, Japan \\
$^{2}$Center for Transdisciplinary, Reserch, Niigata University,
Ikarashi, Niigata 950-2181, Japan \\}

\abst{The electronic states of the FeAs plane in iron-based
superconductors are investigated on the basis of the two-dimensional
16-band d-p model, where the tight-binding parameters are determined so
as to fit the band structure obtained by the density functional
calculation for LaFeAsO. The model includes the
Coulomb interaction on a Fe site: the intra- and inter-orbital direct
terms $U$ and $U'$, the exchange coupling $J$ and the pair-transfer
$J'$. Within the random phase approximation (RPA), we discuss the
pairing symmetry of possible superconducting states including $s$-wave
and $d$-wave pairing on the $U'$-$J$ plane.}

\kword{iron-based superconductors, 16-band d-p model, pairing symmetry, RPA}

\begin{document}
\maketitle
The newly discovered iron-based superconductors\cite{kamihara} RFeAsO$_{1-x}$F$_x$ (R=Rare Earth) with a transition temperature up
to $T_c=55{\rm K}$\cite{ren_3} have attracted much
attention. The mechanism
of the superconductivity is one
of the most significant issues. Several theoretical approaches have been
done on simplified multi-orbital Hubbard models within weak coupling
approaches\cite{kuroki,nomura_1}. The details of the band structure and the Fermi
surface are crucial for determining the pairing symmetry. Therefore, we
employ a realistic model which includes both the Fe 3d orbitals
and the As 4p orbitals, so called d-p model.

 We perform the density functional calculation for LaFeAsO  with the
 generalized gradient approximation of Perdew, Burke and Ernzerhof\cite{perdew} by using the WIEN2k
package\cite{blaha}, where the lattice parameters ($a=4.03268$\AA, $c=8.74111$\AA) and
the internal coordinates ($z_{La}=0.14134$, $z_{As}=0.65166$) are
experimentally determined\cite{nomura}. Considering that there are two distinct Fe and As
sites in the crystallographic unit cell, we
then derive the two-dimensional 16-band d-p model\cite{cvetkovic,yamakawa}, where 3d orbitals ($d_{3z^2-r^2}$, $d_{x^2-y^2}$, $d_{xy}$, $d_{yz}$, $d_{zx}$) of two Fe
atoms (Fe$_1$=$A$, Fe$_2$=$B$) and 4p orbitals ($p_{x}$, $p_{y}$, $p_{z}$) of two As atoms are
explicitly included. We note that
$x, y$ axes are rotated by 45 degrees from the direction along Fe-Fe
bonds. The model is given by the following Hamiltonian,
\begin{eqnarray}
H&=&H_0+H_{\mathrm{int}},\\
H_0&=&\sum_{i,\ell,\sigma}\varepsilon^d_{\ell}d^{\dag}_{i\ell\sigma}d_{i\ell\sigma}+\sum_{i,m,\sigma}\varepsilon^p_{m}p^{\dag}_{im\sigma}p_{im\sigma} \nonumber \\ 
&+&\sum_{i,j,\ell,\ell',\sigma}t^{dd}_{i,j,\ell,\ell'}d^{\dag}_{i\ell\sigma}d_{j\ell'\sigma} \nonumber\\
&+&\sum_{i,j,m,m',\sigma}t^{pp}_{i,j,m,m'}p^{\dag}_{im\sigma}p_{jm'\sigma} \nonumber\\
&+&\sum_{i,j,\ell,m,\sigma}t^{dp}_{i,j,\ell,m}d^{\dag}_{i\ell\sigma}p_{jm\sigma}+h.c. \label{d-p}, 
\end{eqnarray}
where $d_{i\ell\sigma}$ is the annihilation operator for Fe-3d electrons with spin
$\sigma$ in the orbital $\ell$ at the site $i$ and $p_{im\sigma}$ is the annihilation
operator for As-4p electrons with spin
$\sigma$ in the orbital $m$ at the site $i$. In eq. (2), the
transfer integrals $t^{dd}_{i,j,\ell,\ell'}$, $t^{pp}_{i,j,m,m'}$,
$t^{dp}_{i,j,\ell,m}$ and the atomic energies $\varepsilon^d_{\ell}$,
$\varepsilon^p_{m}$ are determined so as to fit both the energy and the weights of orbitals for each band obtained from the
tight-binding approximation (d-p model) to
those from the density functional calculation. 
The doping $x$
corresponds to the number of electrons per unit cell $n=24+2x$ in the present
model. 
\begin{figure}[b]
\begin{center}
\includegraphics[width=5cm]{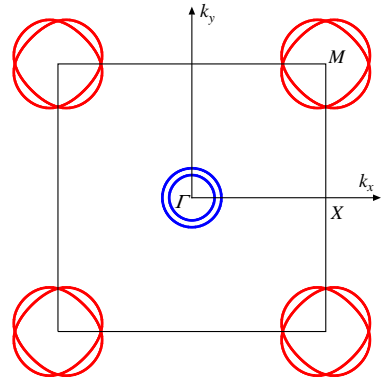}
\end{center}
\caption{Fermi surface obtained from the d-p model eq. (\ref{d-p}) for $x=0.1$ \label{FS}}
\end{figure}

In eq. (1), $H_{\mathrm{int}}$
is the on-site Coulomb interaction in Fe-3d orbitals and
includes the intra-orbital (inter-orbital) Coulomb interaction $U$ ($U'$), 
the Hund's rule coupling $J$ and the pair-transfer interaction $J'$.
In the weak coupling regime, the superconducting gap equation is given by 
\begin{eqnarray}
&&\lambda\Delta^{\alpha\beta}_{\ell\ell'}({\bf k})=\frac{1}{N}\sum_{{\bf k}'}\sum_{\ell_1\ell_2\ell_3\ell_4}\sum_{\alpha',\beta'}\sum_{\mu,\nu}\ \ \ \ \ \ \ \ \ \ \ \ \ \ \nonumber\\
&\times&\frac{f(\varepsilon_{-{\bf k}',\mu})+f(\varepsilon_{{\bf k}',\nu})-1}{\varepsilon_{-{\bf k}',\mu}+\varepsilon_{{\bf k}',\nu}}V^{\alpha,\beta}_{\ell\ell_1,\ell_2\ell'}({\bf k}-{\bf
 k}'){\Delta}^{\alpha'\beta'}_{\ell_3\ell_4}({\bf k}')\nonumber\\
&\times&u^{\alpha'}_{\ell_3,\mu}(-{\bf k}'){u^{\alpha}_{\ell_1,\mu}(-{\bf k}')}^*u^{\beta'}_{\ell_4,\nu}({\bf k}'){u^{\beta}_{\ell_2,\nu}({\bf k}')}^*
\end{eqnarray}
where $\mu$, $\nu$ (=1-16) are band indexes, $\alpha$, $\beta$ ($=$$A,B$) represent two
Fe sites,
$u^{\alpha}_{\ell,\mu}(\mathbf{k})$ is the eigenvector which
diagonalizes $H_0$, $\varepsilon_{\mathbf{k},\mu}$ is the energy of band
$\mu$ with wave vector $\mathbf{k}$, $f(\varepsilon)$ is the Fermi
distribution function. Here, we approximate the effective pairing
interaction
$V^{\alpha,\beta}_{\ell_1\ell_2,\ell_3\ell_4}(\mathbf{q})$
within the RPA\cite{takimoto} and numerically solve the equation (3) to
obtain the gap function $\Delta^{\alpha\beta}_{\ell\ell'}(\mathbf{k})$
with eigenvalue $\lambda$. At $T=T_c$, the largest eigenvalue $\lambda$ becomes unity. We
use $32\times32$ $\mathbf{k}$ points for numerical calculations.

Figure \ref{FS} shows the Fermi surface for the d-p model at $x=0.1$, where
we can see nearly circular hole pockets around the $\Gamma$
point ($\mathbf{k}=(0,0)$) and elliptical electron pockets around the
$M$ point ($\mathbf{k}=(\pi,\pi)$). 
\begin{figure}
\begin{center}
\includegraphics[width=6cm]{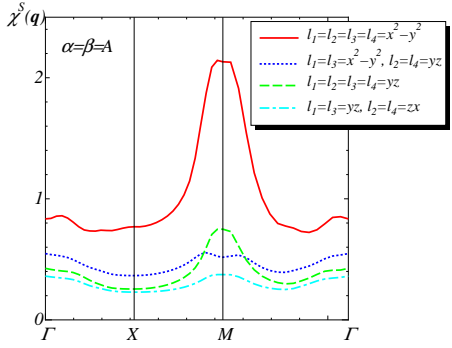}
\end{center}
\caption{Several components of the spin susceptibility  obtained from
 the RPA $\chi^{s~\alpha,\beta}_{\ell_1\ell_2,\ell_3\ell_4}(\mathbf{q})$ for $U=1.5\mathrm{eV}$, $U'=1.0\mathrm{eV}$, $J=J'=0.25\mathrm{eV}$, $T=0.02\mathrm{eV}$ and $x=0.1$. \label{chi}}
\end{figure}
\begin{figure}[t]
\begin{center}
\hspace{-2.5cm}
\begin{minipage}{30mm}
\begin{center}
\includegraphics[width=6.5cm]{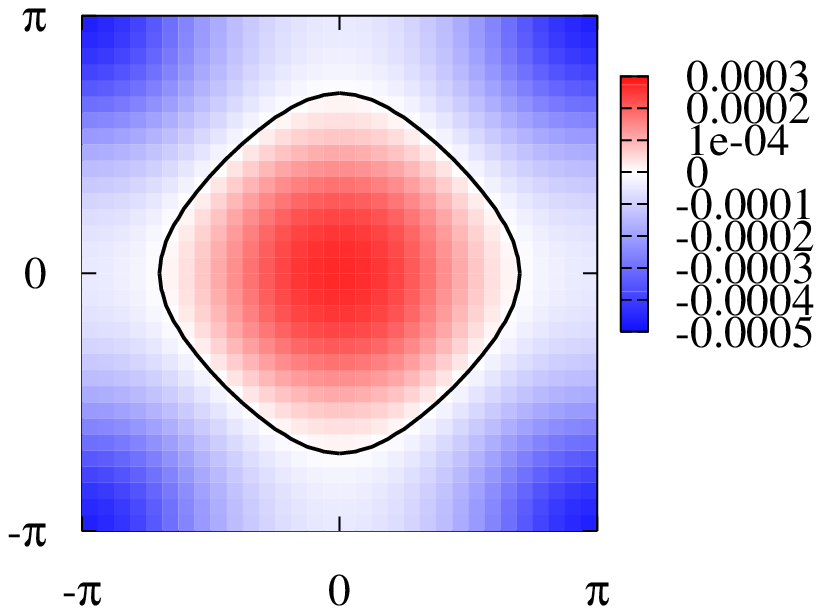}
\end{center}
\end{minipage}
\hspace{3.5cm}
\begin{minipage}{30mm}
\begin{center}
\includegraphics[width=4.1cm]{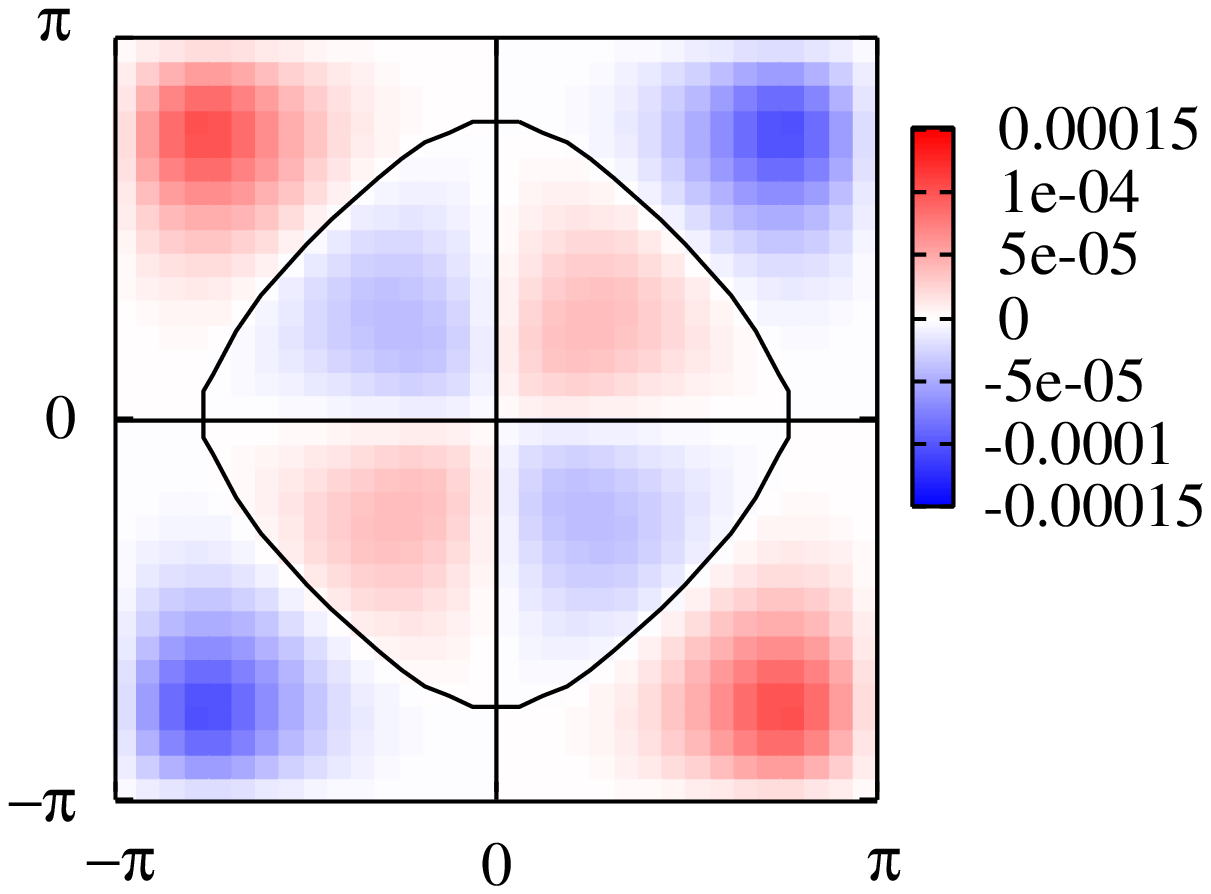}
\end{center}
\end{minipage}
\caption{The gap function
 $\Delta^{AA}_{x^2-y^2,x^2-y^2}(\mathbf{k})$ for
 extended $s$-wave (left) with $\lambda=0.976$ and for $d_{xy}$-wave (right) with
 $\lambda=0.836$
for the same parameters in Fig. \ref{chi}. The solid line
 represents the node of the gap function. \label{gap}}
\end{center}
\end{figure}
The spin
susceptibility 
$\chi^{s~\alpha,\beta}_{\ell_1\ell_2,\ell_3\ell_4}(\mathbf{q})$ is given
by $50\times50$ matrix and is calculated within the RPA as shown in
Fig. \ref{chi}. The peak around
the $\Gamma$ point is due to the effect of the nesting between the
electron (hole) pockets, while the peak around the $M$ point is due to that between the hole pockets and the electron
pockets. We note that the $x^2-y^2$ component becomes dominant around
the $M$ point (see Fig. \ref{chi}).

As the enhanced spin susceptibility contributes to the effective interaction
within the RPA\cite{takimoto}, the gap function for the $x^2-y^2$ component
$\Delta^{AA}_{x^2-y^2,x^2-y^2}(\mathbf{k})$ becomes dominant. The
obtained gap
functions  with two largest eigenvalues $\lambda$ are shown in Fig. \ref{gap}. For
$U=1.5\mathrm{eV}$, $U'=1.0\mathrm{eV}$, $J=J'=0.25\mathrm{eV}$,
$T=0.02\mathrm{eV}$ and $x=0.1$, 
the gap function with the largest $\lambda$ is extended $s$-wave symmetry and changes the sign between the hole
pockets and the electron pockets, while that with second largest $\lambda$ is $d_{xy}$-wave symmetry, as shown in
Fig. \ref{gap}. On the other hand, for $U=1.68\mathrm{eV}$,
$U'=1.4\mathrm{eV}$, $J=J'=0.14\mathrm{eV}$, $T=0.02\mathrm{eV}$ and
$x=0.1$, the gap function with largest $\lambda$ is $d_{xy}$-wave ($\lambda=0.978$),
while that with second largest $\lambda$ is extended $s$-wave ($\lambda=0.967$). The detailed phase
diagram of the pairing symmetry will be shown in the subsequent paper\cite{yamakawa}.

In summary, we investigated the pairing symmetry of the two-dimensional
16-band d-p model by using the
RPA. For a larger value of $J/U'$, the most favorable pairing is
extended $s$-wave symmetry whose order parameter changes
its sign between the hole pockets and the electron pockets, while for a
smaller value of $J/U'$, it is $d_{xy}$-wave symmetry. According to
the recent experiment of very weak $T_c$-suppression by Co-impurities\cite{kawabata}, we suppose that the $d_{xy}$-wave pairing is
suppressed by pair breaking effect
and the extended $s$-wave paring is realized in real materials.

The authors thank M. Sato, H. Kontani and K. Kuroki for useful
comments and discussions. This work was partially supported by the
Grant-in-Aid for Scientific Research from the Ministry of Education,
Culture, Sports, Science and Technology.

\end{document}